\documentclass[aps,prl,twocolumn,reprint,groupedaddress,superscriptaddress]{revtex4-1}

\usepackage{graphicx}
\usepackage{amsmath}

\begin{document}


\title{Triply magic conditions for microwave transitions of optically trapped alkali-metal atoms}


\author{Gang Li}
\email[]{gangli@sxu.edu.cn}
\author{Yali Tian}
\author{Wei Wu}
\author{Shaokang Li}
\author{Xiangyan Li}
\author{Yanxin Liu}
\author{Pengfei Zhang}
\author{Tiancai Zhang}
\email[]{tczhang@sxu.edu.cn}
\affiliation{State Key Laboratory of Quantum Optics and Quantum Optics Devices,
and Institute of Opto-Electronics, Shanxi University, Taiyuan 030006, China}
\affiliation{Collaborative Innovation Center of Extreme Optics, Shanxi University, Taiyuan 030006, China}


\begin{abstract}
We report the finding of ``triply magic'' conditions (the doubly magic frequency-intensity conditions of an optical dipole trap plus the magic magnetic field) for the microwave transitions of optically trapped alkali-metal atoms. The differential light shift (DLS) induced by a degenerate two-photon process is adopted to compensate a DLS associated with the one-photon process. Thus, doubly magic conditions for the intensity and frequency of the optical trap beam can be found. Moreover, the DLS decouples from the magnetic field in a linearly polarized optical dipole trap, so that the magic condition of the magnetic field can be applied independently. Therefore, the ``triply magic'' conditions can be realized simultaneously. We also experimentally demonstrate the doubly magic frequency-intensity conditions as well as the independence of the magnetic field. When the triply magic conditions are fulfilled, the inhomogeneous and homogeneous decoherences for the optically trapped atom will be dramatically suppressed, and the coherence time can be extended significantly.
\end{abstract}



\maketitle

A system of optically trapped cold neutral atoms, especially alkali-metal atoms, is one of the most important testbeds in modern physics. It plays important roles in many research fields, such as quantum simulation \cite{Bloch2008}, quantum metrology \cite{Derevianko2011}, quantum information processing \cite{Reiserer2015}, and quantum computation \cite{Saffman2010}. Most applications rely on the coherence between two ground hyperfine states, where the transition frequency lies in the microwave regime. The decoherence comes from the coupling of the atomic states to the optical dipole trap (ODT) beam and the magnetic field. The main decoherence is due to the inhomogeneous dephasing associated with atomic motion in the ODT \cite{Kuhr2003, Kuhr2005}, in which the fluctuation of the differential light shift (DLS) between the two ground states depends on the kinetic motion of the atom and the local trap intensity. A typical coherence time $T_2$ for atoms trapped in a red-detuned ODT is usually on the 1--100 ms level \cite{Yavuz2006}. Compared to a red-detuned ODT, a blue-detuned ODT has weaker inhomogeneous decoherence \cite{Li2012, Tian2019}. By applying a series of $\pi$-pulses, the inhomogeneous dephasing process can be technically recovered, and $T_2$ can thus be greatly elongated \cite{Dudin2013, Yu2013}; $T_2$ is finally limited by other weaker homogeneous dephasing factors, such as ODT power fluctuation, ODT beam pointing noise, and variation in the magnetic field.

To physically suppress both inhomogeneous and homogeneous dephasings and extend $T_2$, ``magic'' trapping conditions, such as magic wavelength, magic polarization, magic intensity, and magic magnetic field, have been proposed and investigated extensively in recent years \cite{Jai2007, Flambaum2008, Derevianko2010, Derevianko2011, Carr2016, Lundblad2010, Dudin2010, Radnaev2010, Chicireanu2011, Kim2013, Sarkany2014, Kazakov2015, Yang2016}. By deliberately arranging the wavelength, polarization, and intensity of ODT beam, or the magnetic field, some of the microwave transition can be either immune to the fluctuations of the trapping beam or to the magnetic field alone. This single magic condition helps to increase $T_2$ for the corresponding microwave transition \cite{Lundblad2010, Dudin2010, Chicireanu2011, Kim2013, Sarkany2014, Kazakov2015, Yang2016}. To further extend time $T_2$, ``doubly magic'' conditions, where the transition is immune to fluctuations in both the ODT beam and the magnetic field, are preferred. A set of ``doubly magic'' conditions (the magic wavelength of the ODT plus the magic magnetic field) were proposed for the multiphoton ($m_F=-n \leftrightarrow m'_F=n$ between two ground hyperfine Zeeman states with $n \ge 2$) microwave transitions of cesium by matching the polarization, wavelength, and direction of the ODT beam to the magnetic field \cite{Flambaum2008, Derevianko2010}. The most recent investigation showed that the ``doubly magic'' conditions (the magic intensity of the ODT plus the magic magnetic field) can also be found with bichromatic ODT beams by taking into account the hyperpolarizability and the coupling between the circularly polarized ODT beam and magnetic field \cite{Carr2016}. However, in these ``doubly magic'' conditions, either the multiphoton microwave transitions or bichromatic ODT beams are prerequisite and both are not favorable for the experiment. 

Here, we theoretically propose and experimentally demonstrate a new scheme to realize ``triply magic'' conditions (the doubly magic frequency-intensity conditions of the ODT beam plus the magic magnetic condition) for the microwave transition ($m_F=-n \leftrightarrow m'_F=n$ with $n \ge 0$) of an alkali-metal atom with a monochromatic ODT beam. Moreover, the doubly magic frequency-intensity conditions of the ODT beam can be found for any microwave transition. The magic magnetic condition does not couple to the ODT beam with linear polarization and can be applied independently. When the ``triply magic'' conditions are met, $T_2$ can be substantially prolonged with current experimental capabilities. The ultimate $T_2$ is then limited by the Raman scattering of the ODT photons and storage time of the trapped atom. 

\begin{figure}
\includegraphics[width=8.6 cm]{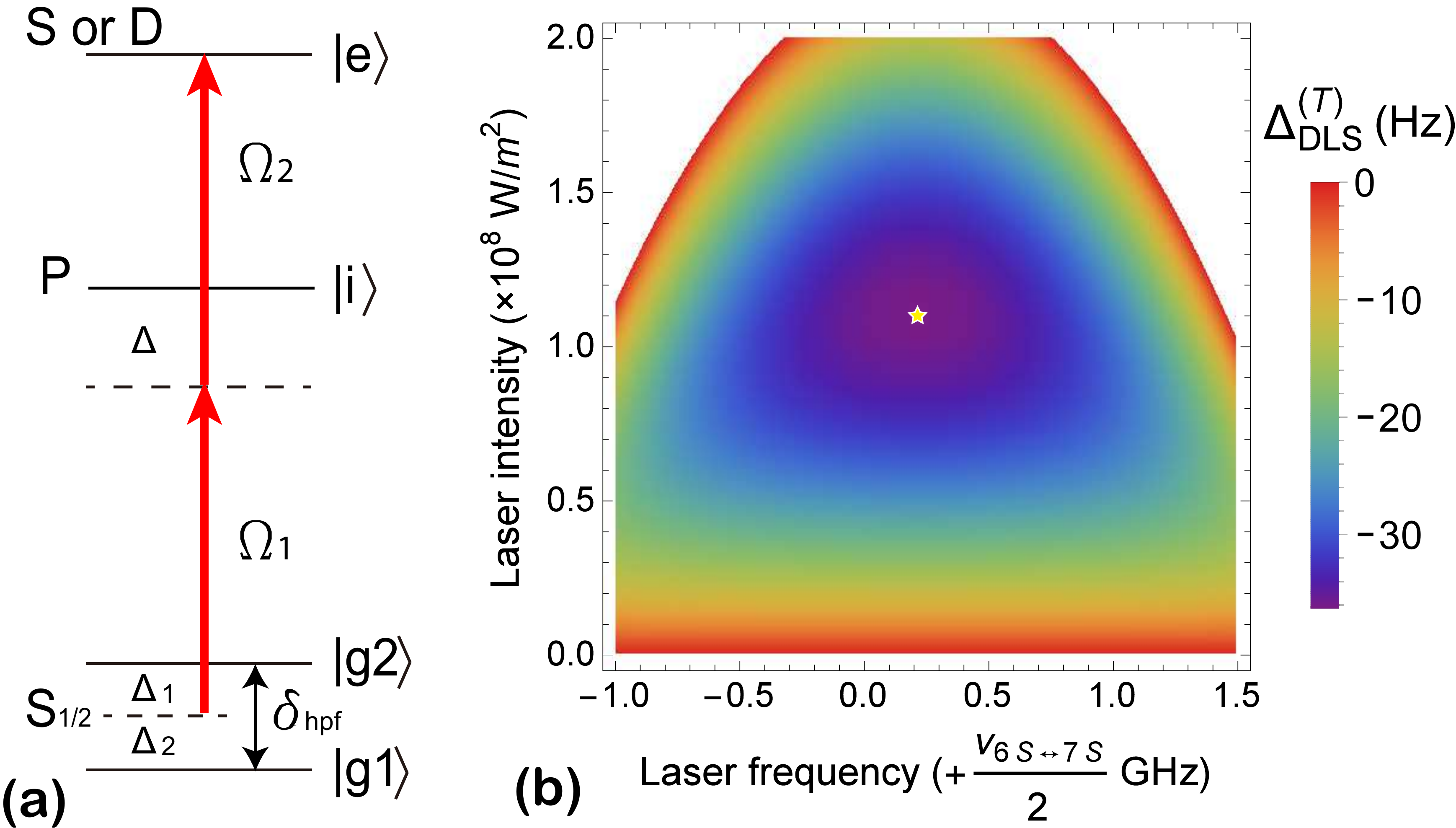}
\caption{\label{fig1} (a) Energy level scheme for atoms with the OPP and the additional TPP. (b) A 2D plot of total DLS between cesium clock states $|6 \text{S}_{1/2},F=3,m_F=0\rangle$ and $|6 \text{S}_{1/2},F=4,m_F=0\rangle$ for the ODT beam frequency and intensity, with 7S$_{1/2}$ state as the TPP excited state. The ``star'' marks the doubly magic frequency-intensity point. }
\end{figure}

The general idea of our method is to use the DLS induced by a degenerate two-photon process (TPP) to compensate the DLS associated with a one-photon process (OPP). Both the TPP and OPP are induced by the same monochromatic ODT light field. Figure \ref{fig1}(a) shows a simple model with two ground states $|g1\rangle$ and $|g2\rangle$, one intermediate state $|i\rangle$, and one excited state $|e\rangle$. $|g1\rangle$ and $|g2\rangle$ are the two Zeeman states in the hyperfine doublet of the ground S state for an alkali-metal atom. $|e\rangle$ is a hyperfine state in the higher S or D state, and $|i\rangle$ is a hyperfine state in the intermediate P state. The OPP couples $|g1(2)\rangle$ and $|i\rangle$, and the TPP couples $|g1(2)\rangle$ and $|e\rangle$ via $|i\rangle$. In this letter, we only consider a linearly polarized ODT with the polarization parallel to the direction of the quantization magnetic field. The induced DLS then decouples from the magnetic field and can be applied independently.

Let us first consider the DLS associated with the OPP [see lower part of Fig. \ref{fig1}(a)]. Regardless of whether the trap frequency is red-detuned or blue-detuned to the atomic transitions $|g1(2)\rangle\leftrightarrow |i\rangle$, the DLS between two ground states,
\begin{equation}\label{eq1}
\delta^\text{(1p)}_\text{DLS}=\delta^\text{(1p)}_2-\delta^\text{(1p)}_1=\frac{-\delta_\text{hpf}}{4 \Delta^2} \Omega_1^2
\end{equation}
is always negative due to the negative differential frequency detuning $-\delta_\text{hpf}$ \cite{Rosenbusch2009}. Here, $\delta^\text{(1p)}_{1(2)}$ is the light shift of state $|g1(2)\rangle$, $\delta_\text{hpf}$ is the hyperfine splitting between two ground hyperfine states, and $\Omega_1=|\langle i|\hat{d}|g1(2)\rangle| E/\hbar$ is the Rabi frequency of the ODT beam, where $E$ is the electric strength of the trap beam. Therefore, the amount of DLS is proportional to the light intensity $I$ and has the same spatial distribution. The fluctuations in the DLS caused by the thermal motion of the atom and the noise associated with the ODT field result in inhomogeneous and homogeneous dephasings between the two ground states. 

To suppress the fluctuations in the DLS, we consider an additional DLS associated with the TPP in Fig. \ref{fig1}(a). The TPP frequency detunings for transitions from $|g1\rangle$ and $|g2\rangle$ to excited state $|e\rangle$ are $\Delta_1$ and $\Delta_2=\Delta_1-\delta_\text{hpf}$, respectively. The one-photon detuning from $|g2\rangle$ to intermediate state $|i\rangle$ is then $\Delta=(\omega_\text{ei}-\omega_\text{ig2}+\Delta_1)/2 \approx (\omega_\text{ei}-\omega_\text{ig2})/2$ for $\Delta_1 \ll (\omega_\text{ei}-\omega_\text{ig2})$. $\omega_\text{ei}$ and $\omega_\text{ig2}$ are the resonant frequencies for transitions $|e\rangle \leftrightarrow|i\rangle$ and $|i\rangle \leftrightarrow|g2\rangle$, respectively. Thus, the effective Rabi frequency for the TPP is $\Omega_\text{TPP}=\frac{\Omega_1\Omega_2}{\Delta}$, where $\Omega_1$ and $\Omega_2$ are the one-photon Rabi frequencies, which couple the lower and higher parts of the two-photon transition. The TPP-induced light shifts for $|g1\rangle$ and $|g2\rangle$ are $\delta^\text{(2p)}_1=\frac{\Omega_\text{TPP}^2}{4 \Delta_1}$ and $\delta^\text{(2p)}_2=\frac{\Omega_\text{TPP}^2}{4 \Delta_2}$, respectively. The DLS between $|g2\rangle$ and $|g1\rangle$ associated with the TPP is the subtraction
\begin{equation}\label{eq2}
\delta^\text{(2p)}_\text{DLS}=\delta^\text{(2p)}_2-\delta^\text{(2p)}_1=\frac{\delta_\text{hpf}}{4 \Delta^2 \Delta_1(\delta_\text{hpf}-\Delta_1)} \Omega_1^2\Omega_2^2.
\end{equation}

The Rabi frequencies $\Omega_1=|\langle i|\hat{d}|g1(2)\rangle| \sqrt{\frac{2I}{c \epsilon_0}} /\hbar$ and $\Omega_2=|\langle e|\hat{d}|i \rangle| \sqrt{\frac{2I}{c \epsilon_0}} /\hbar$, where $c$ and $\epsilon_0$ are the speed of light and vacuum permittivity, respectively. The total DLS $\Delta^\text{(T)}_\text{DLS}$ is then the sum of Eqs. (\ref{eq1}) and (\ref{eq2}). Therefore, we determine that the first-order derivative of the total DLS with respect to both $\Delta_1$ and $I$ vanishes at the points of
\begin{equation}\label{eq3}
\Delta_1=\delta_\text{hpf}/2
\end{equation}
and
\begin{equation}\label{eq4}
I_0= \frac{\delta_\text{hpf}^2}{8}  \frac{\hbar^2 c \epsilon_0}{|\langle e|\hat{d}|i\rangle|^2},
\end{equation}
where the variance of the total DLS depends on the fluctuations of $\Delta_1$ and $I$ only on the second order. Eqs. (3) and (4) give the magic frequency and intensity, respectively. However, these doubly magic conditions are obtained from a simple model in which only one intermediate hyperfine state and one excited hyperfine state are involved. To apply this model to a real atom, all the hyperfine states in the intermediate P states and higher excited S or D state should be considered. The DLS then takes the sum. The details of the theoretical framework are given in the Supplemental Materials (SM) \cite{N30} .

To illustrate our method, we calculate the total DLS between cesium clock states $|6\text{S}_{1/2},F=3,m_F=0\rangle$ and $|6\text{S}_{1/2},F=4,m_F=0\rangle$ [we denote them as (0,0) hereafter] in a 1079-nm ODT, in which the TPP couples the two ground states in 6S$_{1/2}$ to hyperfine states in 7S$_{1/2}$. Since there are two hyperfine states with $F''=3$ and 4 in the 7S$_{1/2}$ state, only two TPP transition lines ($F=3 \leftrightarrow F''=3$ and $F=4 \leftrightarrow F''=4$) are allowed with the linearly polarized beam. The calculation of the DLS for the ground states needs to sum the contributions from both transition lines via all intermediate states in the hyperfine multiplet of 6P$_{1/2, 3/2}$. It should be pointed out that the TPP selection rules of $\Delta F=0$ and $\Delta m_F=0$ \cite{Cagnac1973, Antypas2014} simplify the calculation significantly in our special case (single beam, degenerate two-photon excitation). If one deviates from these assumptions, then the calculation becomes more complicated and will change the results. The DLS of the OPP is calculated by taking into account the states $n\text{P}_{1/2, 3/2}$ with principle quantum number $n=6,7,...,14$. The dependence of the total DLS on the frequency and intensity of the ODT is shown in Fig. \ref{fig1}(b). The frequency is given with the reference being the half-distance from 6S to 7S. Obviously, there is a DLS minimum at $\nu_0=\nu_{6S \leftrightarrow 7S}/2+0.219$ GHz and $I_0=1.11\times 10^8$ W/m$^2$. This point is marked by a yellow star in Fig. \ref{fig1}(b). The corresponding trap depth is $U_T=-22.5$ $\mu$K, which can be constructed by a strongly focused laser beam with a 2.5-$\mu$m waist and 1.1-$m$W power. At the points of DLS minima, the first-order dependences of the DLS on both the trap frequency and intensity vanish. The residual coefficients for the second-order DLS with respect to the ODT frequency and intensity are $k_\nu=\partial^2 \Delta^\text{(T)}_\text{DLS}/\partial \nu^2|_{(I_0, \nu_0)} = 23.6 \times 10^{-18}$ Hz$^{-1}$ and $k_I=\partial^2 \Delta^\text{(T)}_\text{DLS}/\partial I^2 |_{(I_0, \nu_0)} = 5.93 \times 10^{-15}$ Hz$\cdot $m$^4/$W$^2$. These results are summarized in the third row of Table \ref{tab1}.

\begin{figure}
\includegraphics[width=8.6 cm]{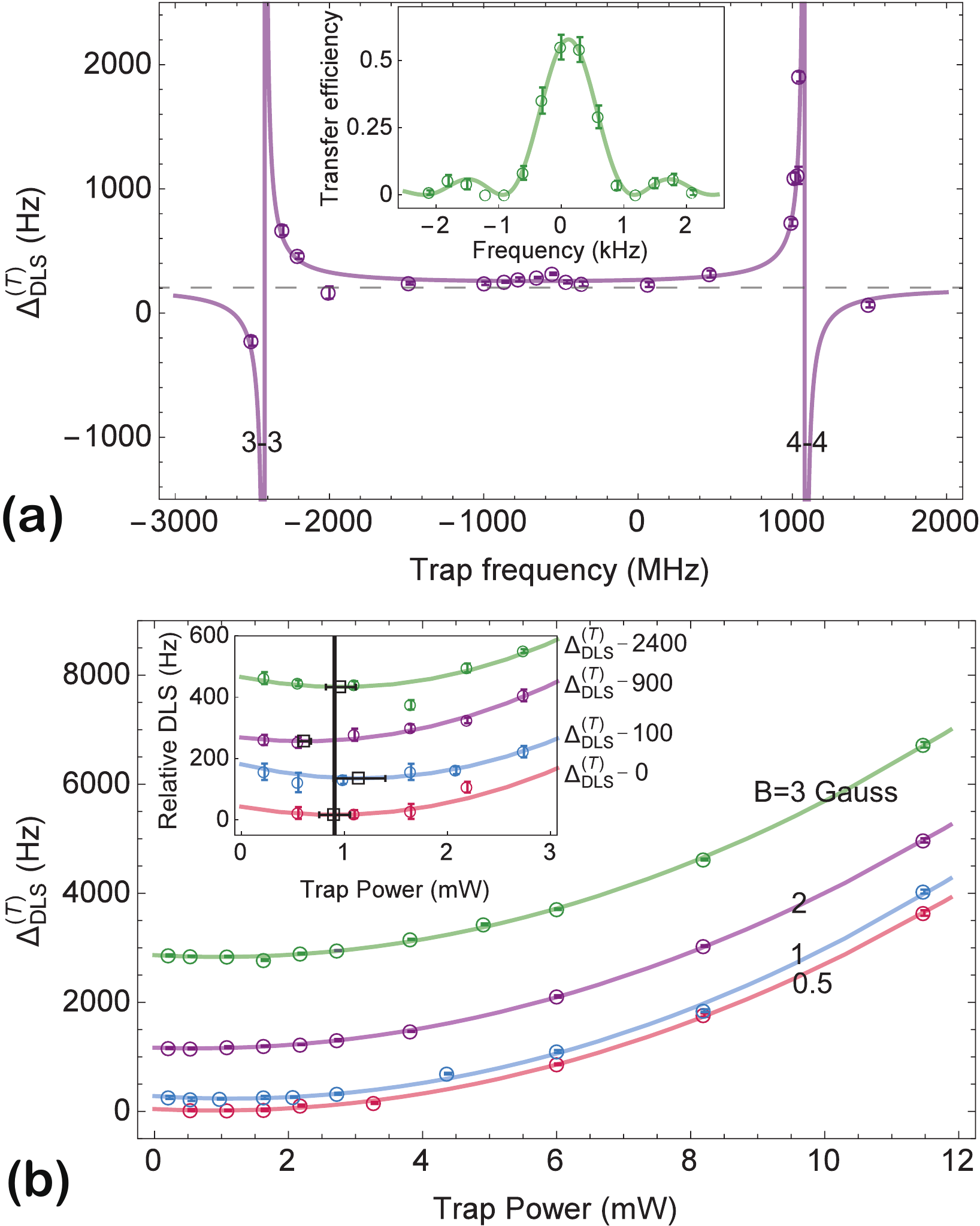}
\caption{\label{fig2}  The experimental results for the total DLS of cesium (0,0) states versus laser frequency (a) and laser power (b) in a 1079-nm optical trap. 3--3 and 4--4 in (a) indicate the two TPP transition lines of $F=3 \leftrightarrow F''=3$ and $F=4 \leftrightarrow F''=4$ between 6S and 7S. The trap power is approximately 1.4 m$W$, and the strength of the magnetic field is approximately 1 Gauss for the measurements in (a). The trap frequency is locked at the frequency point of $-670$ MHz [the frequency at the right middle of the two transition lines in (a)] for the measurements in (b). The inset of (a) displays a typical spectrum for the (0,0) transition. The inset of (b) shows the zoomed dependence of total DLS on laser power around the DLS minimum with certain DLS shifts for every set of data. The black open squares are the magic powers under different B-fields, and the vertical black line is the mean value.}
\end{figure}

We then experimentally demonstrate the doubly magic frequency-intensity conditions on a single cesium atom trapped in a microsized 1079-nm trap. The ODT is constructed by a strongly focused laser beam with linear polarization parallel to the direction of the magnetic field. The details of the experimental layout can be found in \cite{Tian20191}. The DLS of cesium (0,0) states is obtained by fitting the transition spectrum [inset of Fig. \ref{fig2}(a)] \cite{N10} at the specific trap frequency, trap intensity, and magnetic field.

Figure \ref{fig2}(a) shows the measurements of the total DLS when the frequency of the trap beam scans point by point over the two transition lines $F=3 \leftrightarrow F''=3$ and $F=4 \leftrightarrow F''=4$, which are located at frequencies of 1.08 GHz and $-2.42$ GHz in the figure, respectively. The minima of the total DLS occurs in the right middle of the two transitions, and the data fitting gives a residual coefficient of $33.7(1.7) \times 10^{-18}$ Hz$^{-1}$. The trap power we used for these measurements is 1.4 mW. By taking the 2.5-$\mu$m beam size into account, the light intensity is $1.54 \times 10^8$ W/m$^2$ at the trap bottom, which is higher than the magic intensity. At this light intensity, the theoretical second-order coefficient of the DLS on the ODT frequency is $39.5 \times 10^{-18}$ Hz$^{-1}$, which agrees well with the experimental results.

Figure \ref{fig2}(b) shows the measurements of the total DLS when the power of the 1079-nm ODT scans point by point from low to high under different magnetic field strengths (B-field). The total DLS depends on the trap power quadratically. The fitting of the experimental results by a quadratic function gives the magic power under different B-fields, which are shown in the inset of Fig. \ref{fig2}(b) with back open squares. As expected, the magic power is independent of the B-field. The average magic power is 0.91(0.15) mW, and the residual coefficient for the second-order derivation of the DLS to the ODT power is $67.4(2.2)$ Hz/mW$^2$. If we consider that the ODT has a waist of approximately 2.5 $\mu$m, the deduced magic intensity and residual second-order coefficient are $1.0(0.2)\times 10^8$ W/m$^2$ and $6.2(0.2) \times 10^{-15}$ Hz$\cdot $m$^4/$W$^2$, respectively. They are in good agreement with the theoretical calculations.

\begin{table*}
\caption{\label{tab1} Doubly magic frequency-intensity conditions of the ODT beam for cesium by 2-photon coupling to a higher energy level. GS means ground states with (m,n) representing ($|6S_{1/2},F=3,m_F=\text{m}\rangle$, $|6S_{1/2},F=4,m_F=\text{n}\rangle$). ES means the TPP excited state. $\lambda$, $\Delta \nu_0=\nu_0-\nu_\text{fine}/2$, and $I_0$ are the wavelength, relative frequency, and intensity of the trap beam at the magic point, respectively. $\nu_0$ and $\nu_\text{fine}$ are the laser frequency and frequency distance between the fine structures of ground state and higher excited state. $U_T$ is the trap potential. $k_\nu=\partial^2 \Delta^\text{(T)}_\text{DLS}/\partial \nu^2|_{(I_0, \nu_0)} $ and $k_I=\partial^2 \Delta^\text{(T)}_\text{DLS}/\partial I^2|_{(I_0, \nu_0)}$ are the residual coefficients for the second-order derivation of the total DLS over the trap beam frequency and intensity. $\Gamma^\text{(1p)}_\text{RS}$ and $\Gamma^\text{(2p)}_\text{S}$ are the 1-photon Raman scattering rate and 2-photon scattering rate, respectively. The wavelengths and transition matrices for the calculation are from \cite{Sansonetti2009} and \cite{Safronova2016}.}
\begin{ruledtabular}
\begin{tabular}{ccccccccccc}
GS & ES & $\lambda$& $\Delta \nu_0$ & $I_0$ & $U_T$ & $k_\nu$ & $k_I$ & $\Gamma^\text{(1p)}_\text{RS}$ &$\Gamma^\text{(2p)}_\text{S}$\\
& & (nm) & (GHz) & ($\text{GW}/\text{m}^2$)& ($\mu$K) & ($\times 10^{-18}$ Hz$^{-1}$) & (fHz$\cdot$ m$^4$/W$^2$) & (Hz) & (Hz)\\
\hline
(0,0) & 5D$_{3/2}$ & 1379 & 0.281 & 0.0195 & $-1.8$ & 0.587 & 8.76 & $2.0\times 10^{-6}$ & $3.0\times 10^{-6}$ \\
(0,0) & 5D$_{5/2}$ & 1370 & 0.290 & 0.0146 & $-1.4$ & 0.462 & 12 & $1.6\times 10^{-6}$ & $1.7\times 10^{-5}$ \\
(0,0) & 7S$_{1/2}$ & 1079 & 0.219 & 0.111 & $-22.5$ & 23.6 & 5.93 & $3.4\times 10^{-4}$ & $1.1\times 10^{-2}$ \\
($-1$,1)& 7S$_{1/2}$ & 1079 & 0.219 & 0.110 & $-22.3$ & 23.3 & 5.93 & $3.4\times 10^{-4}$ & $1.1\times 10^{-2}$ \\
(3,4)& 7S$_{1/2}$ & 1079 & 0.219 & 0.102 & $-20.7$ & 20.0 & 5.93 & $3.1\times 10^{-4}$ & $9.3\times 10^{-3}$\\
(0,0) & 7D$_{3/2}$ & 767.8 & 0.286 & 0.429 & $154$ & 125 & 3.63 & $1.1\times 10^{-2}$ & $1.0\times 10^{-5}$ \\
(0,0) & 7D$_{5/2}$ & 767.2 & 0.287 & 0.155 & $55.1$ & 44.9 & 9.89 & $4.0\times 10^{-3}$ & $2.7\times 10^{-5}$ \\
\end{tabular}
\end{ruledtabular}
\end{table*}

In addition to the above magic frequency-intensity conditions of the 1079-nm ODT for (0,0), in principle, they can also be found for any microwave transitions. For example, the states $(-1,1)$ have the magic condition of the magnetic field and have great potential for obtaining a very long coherence time when all three magic conditions are met. Moreover, in many quantum manipulation protocols, for example, photon-atom logic \cite{Reiserer2014, Duan2004} and digital atom interferometer \cite{Steffen2012}, the atomic qubit is encoded in a pair of hyperfine ground states in which the atom interacts with the optical field differently. The states (3,4) or $(-3,-4)$ in cesium are often adopted for this purpose. For the sake of these experiments, the frequency-intensity magic conditions for $(-1,1)$ and (3,4) are also calculated and listed in Table \ref{tab1}. 

In addition to the 7S$_{1/2}$ state coupled by the 1079-nm laser, there are many other states, such as 5D$_{1/2}$, 5D$_{3/2}$, 6D$_{1/2}$, 6D$_{3/2}$, 8S$_{1/2}$, 7D$_{1/2}$, and 7D$_{3/2}$, that can be used as the higher excited states of the TPP for realizing the magic frequency-intensity conditions of cesium. For these states, more intermediate states, such as 7P$_{1/2,3/2}$ and 8P$_{1/2,3/2}$, need to be taken into account. The magic conditions of the ODT with 5D$_{1/2}$, 5D$_{3/2}$, 7D$_{1/2}$, and 7D$_{3/2}$ are also calculated and listed in Table \ref{tab1}. The cases for the states 6D$_{3/2}$, 6D$_{5/2}$, and 8S$_{1/2}$ are absent because of high one- and two-photon scattering rates. 

Since our doubly magic frequency-intensity conditions are independent of the magnetic field, the third magic condition of the magnetic field for states $(-n,n)$ can be applied independently. The magic magnetic condition of cesium (0,0) states calculated by the Breit-Rabi formula is $B_0=0$ Gauss, at which the residual second-order coefficient for the differential Zeeman shift over the magnetic field strength is $k_M=\partial^2 \Delta_\text{Zeeman}/\partial B^2|_{B_0}=854.9$ Hz$/\text{Gauss}^2$. The magic condition of the magnetic field for $(-1,1)$ is $B_0=1.39$ Gauss, and the residual second-order coefficient is $k_M=801.5$ Hz$/\text{Gauss}^2$. This magic magnetic-field condition plus the magic frequency-intensity conditions makes our scheme have triply magic conditions.

The trap depth for the magic conditions of some red (negative) traps for cesium in Table \ref{tab1} might be too shallow to directly load atoms from a laser-cooled atomic ensemble. We propose using a polychromatic trap to resolve this dilemma. The trap beam is phase modulated with a certain frequency, and the DLS associated with parts of nondegenerate TPPs will compensate the DLS from the degenerate TPP. As a result, the magic trap depth will increase at the cost of relatively higher 1- and 2-photon scattering rates. The details are shown in the SM \cite{N30}.

As an outlook, when the triply magic conditions are fulfilled, the ground states' $T_2$ time would be extended dramatically. The measurable $T_2$ time is limited by the atom-motion-induced inhomogeneous dephasing and the Raman-scattering-induced state lifetime $T_1$. In the 1079-nm magic ODT, a theoretical analysis (see SM \cite{N30}) shows that $T_2=100$ s could be feasible with an atom below the temperature of 0.2 $\mu$K. The one-photon Raman scattering rate $\Gamma^\text{(1p)}_\text{RS}$ can be calculated by the Kramers-Heisenberg formula \cite{Loudon2000, Cline1994}. The two-photon Raman scattering rate $\Gamma^\text{(2p)}_\text{RS}$ can be estimated by the scattering rate of the higher excited state $\Gamma^\text{(2p)}_\text{S}$, which sets an upper limit on the two-photon Raman scattering rate. Both scattering rates for the listed magic conditions in Table \ref{tab1} are calculated and displayed in the same table. Then, the state lifetime $T_1=91$ s, which is mainly determined by two-photon scattering, sets the new limit for $T_2$. Therefore, the  laser frequency, laser power, and magnetic field need to be controlled within the accuracies of $\Delta \nu <\pm 10$ MHz, $\Delta I/I_0 <\pm 6$\%, and $\Delta B <\pm 2$ mGauss, respectively, without significantly impacting $T_2$ \cite{N20}. 

In conclusion, we have presented a new scheme to realize ``triply magic'' trapping conditions for cesium atom microwave transitions (the doubly magic frequency-intensity conditions of the ODT plus the magic magnetic field). The doubly magic frequency-intensity conditions of the ODT can be found for any ground state microwave transitions. Some experimentally favorable magic frequency-intensity conditions with different wavelengths are found. We also experimentally demonstrate the magnetic field-independent magic frequency-intensity conditions with a single cesium atom in a 1079-nm trap, and the results prove the feasibility of the triply magic conditions. In addition to cesium, similar magic conditions can also be found for other atomic species, such as rubidium (see the SM) \cite{N30}.

\begin{acknowledgements}
We thank Dr. Jie Li, Dr. J. Manz, and Arif Kamal for careful reading of the manuscript. This work is supported by the National Key Research and Development Program of China (Grant No. 2017YFA0304502), the National Natural Science Foundation of China (Grant Nos. 11634008, 11674203, 11574187, 11974223, and 11974225), and the Fund for Shanxi "1331 Project" Key Subjects Construction. 

G.L. and Y.T. contributed equally to this work. G.L., P.Z., and T.Z. conceptualized the work and developed the theory; Y.T., W.W., S.L., X.L., and Y.L. conducted the experiments under supervision of G.L. and T.Z.. 
\end{acknowledgements}

\bibliography{MagicCondition}

\begin{thebibliography}{43}
\bibitem [{N10()}]{N10}%
  \BibitemOpen
  \href@noop {} {}\bibinfo {note} {The spectrum is obtained by the following
  experimental sequences: 1) preparing the atom into 6S$_{1/2}$ $F=4, m_F=0$ by
  optical pumping; 2) applying a microwave pulse with frequency $f_{\mu
  \text{w}}$ to transfer the atom in state 6S$_{1/2}$ $F=4, m_F=0$ to state
  6S$_{1/2}$ $F=3, m_F=0$; 3) making the state detection to obtain the transfer
  efficiency; 4) scanning $f_{\mu \text{w}}$ to another value and repeating
  steps 1-3.}\BibitemShut {Stop}%
\bibitem [{N20()}]{N20}%
  \BibitemOpen
  \href@noop {} {}\bibinfo {note} {The tolarances of the noises on laser
  frequency, laser power, and magnetic field are calculated by assuming that
  the $T_2$ time caused by every noise factor is longer than 360
  s.}\BibitemShut {Stop}%
\end{thebibliography}

\pagebreak

\clearpage

\widetext

\renewcommand{\thefigure}{S\arabic{figure}}

\renewcommand{\thetable}{S\arabic{table}}

\section{Supplementary information for: Triply magic conditions for microwave transition of optically trapped alkali-metal atoms}

\subsection*{ Gang Li, Yali Tian, Wei Wu, Shaokang Li, Xiangyan Li, Yanxin Liu, Pengfei Zhang, and Tiancai Zhang}

\subsection*{State Key Laboratory of Quantum Optics and Quantum Optics Devices,
and Institute of Opto-Electronics, Shanxi University, Taiyuan 030006, China\\
Collaborative Innovation Center of Extreme Optics, Shanxi University, Taiyuan 030006, China
}

\affiliation{State Key Laboratory of Quantum Optics and Quantum Optics Devices,
and Institute of Opto-Electronics, Shanxi University, Taiyuan 030006, China}
\affiliation{Collaborative Innovation Center of Extreme Optics, Shanxi University, Taiyuan 030006, China}

\section{calculation of ac Stark energy shifts associated with the one-photon and two-photon processes}

The ac Stark energy shift of the ground state $|g\rangle=|J_g,F_g,m_g\rangle$ due to one-photon coupling to the excited states $|i\rangle=|J_i,F_i,m_i\rangle$ by a linear polarized light field $E(t)=E_0 \cos(\omega t)$ can be calculated by perturbation theory to the first order, and the result is \cite{Cohen1998}
\begin{equation}\label{eq1s}
U^\text{(1p)}_g= \sum_{i} \left( \frac{|\langle i|\mathcal{H}|g\rangle|^2}{E_i-E_g-\hbar \omega}+\frac{|\langle i|\mathcal{H}|g\rangle|^2}{E_i-E_g+\hbar \omega}\right),
\tag{S1}
\end{equation}
where $\mathcal{H}$ is the interaction Hamiltonian; $E_i$, $E_g$, and $\hbar \omega$ are the energies of states $|i\rangle$, $|g\rangle$, and the photon, respectively. The two terms in the parenthesis are the rotating and the counter-rotaing terms. The interaction Hamiltonian can be expressed as $\mathcal{H}= - \hat{d} E_0 /2$ in the rotating frame associated with the light field, where $\hat{d}=-e r$ is the electric dipole operator. Substituting the one photon Rabi frequency $\Omega_{i}= E_0 \langle i|\hat{d}|g\rangle/\hbar$ into Eq.
(\ref{eq1s}), the energy shift can be expressed as
\begin{equation}\label{eq2s}
U^\text{(1p)}_g=\frac{\hbar ^2}{4}\sum_{i}\Omega_{i}^2 \left( \frac{1}{E_i-E_g-\hbar \omega}+\frac{1}{E_i-E_g+\hbar \omega}\right),
\tag{S2}
\end{equation}
where $\langle i |\hat{d}|g \rangle$ is the transition dipole moment.

When a higher excited state $|e\rangle=|J_e F_e m_e\rangle$ and a two-photon process coupling $|g\rangle$ and $|e\rangle$ via $|i\rangle$ are taken into account, the two-photon ac Stark energy shift of the ground state is then
\begin{equation}\label{eq3s}
U^\text{(2p)}_g=\frac{\hbar}{2}\sum_{e}\left(\sqrt{\Omega_{\text{TPP},e}^2+\Delta_e^2}-\Delta_e\right),
\tag{S3}
\end{equation}
where $\Delta_e=(E_e-E_g)/\hbar-2\omega$ is the two photon frequency detuning, and $\Omega_{\text{TPP},e}$ is the effective Rabi frequency of the TPP. Here we omit the counter-rotating wave term due to $\Delta_e \ll (E_e-E_g)/2 \hbar$. If $\Delta_e \gg \Omega_{\text{TPP},e}$, Eq. (\ref{eq3s}) can be approximated by
\begin{equation}\label{eq4s}
U^\text{(2p)}_g=\frac{\hbar}{4}\sum_{e}\frac{\Omega_{\text{TPP},e}^2}{\Delta_e}.
\tag{S4}
\end{equation}
The Rabi frequency of the TPP can be calculated by the perturbation theory to the second order \cite{Cohen1998}. It gives 
\begin{multline}\label{eq5s}
\Omega_{\text{TPP},e}\\
= \sum_{i}\Big[\frac{\langle e|\mathcal{H}_{2}| i \rangle \langle i|\mathcal{H}_{1}|g\rangle}{(E_i-E_g-\hbar \omega_1)/\hbar} +\frac{\langle e|\mathcal{H}_{1}| i\rangle \langle i |\mathcal{H}_{2}| g \rangle}{(E_i-E_g-\hbar \omega_2)/\hbar}\Big],\\
\tag{S5}
\end{multline}
where $\mathcal{H}_{1}$ and $\mathcal{H}_{2}$ are the Hamiltonians that describe the dipole interactions between the atom and two light fields enrolled in the general TPP. $\omega_1$ and $\omega_2$ are the frequencies of these two light fields. In this model, the two light fields are the same, thus  $\mathcal{H}_{1}=\mathcal{H}_{2}=- \hat{d} E_0 /2$ and $\omega_1=\omega_2=\omega$. Therefore, after substituting single photon Rabi frequencies $\Omega_{i,1}=E_0 \langle i|\hat{d}|g\rangle/ \hbar$ and $\Omega_{i,2}=E_0 \langle e|\hat{d}|i\rangle/ \hbar$ in Eq. (\ref{eq5s}), the effective Rabi frequency of the TPP can be calculated through
\begin{equation}\label{eq6s}
\Omega_{\text{TPP},e}= \sum_{i}\frac{\Omega_{i,1} \Omega_{i,2}}{\delta_i},
\tag{S6}
\end{equation}
where $\delta_i=(E_i-E_g-\hbar \omega)/\hbar$ is the single photon detuning between the trap light field and the lower part of one photon transition in the TPP.

The transition dipole moment can be calculated by the 3$j$-symbol, 6$j$-symbol and a reduced dipole matrix element between two fine states:
\begin{equation}\label{eq7s}
  \langle J_j F_j m_j|\hat{d}|J_l F_l m_l\rangle= \langle L_j||\hat{d}||L_l\rangle (-1)^{2 F_j+L_l+I+m_l}\\
  \times \sqrt{(2F_j+1)(2F_i+1)}
  \begin{pmatrix}
  F_j & 1 & F_l\\
  m_j & p & -m_l
  \end{pmatrix}
  \begin{Bmatrix}
  J_l & J_j & 1\\
  F_j & F_l & I
  \end{Bmatrix},
\tag{S7}
\end{equation}
where $l$ and $j$ can be $g$, $i$, and $e$ depending on the transition; $J_{l,j}$, $F_{l,j}$, and $m_{l,j}$ are the orbital, hyperfine, and magnetic quantum numbers of the two atomic states; $I$ is the nuclear spin, and $p$ is the polarization of the light field. In this case, only the linearly polarized light field was used, hence $p=0$. $\langle L_j||\hat{d}||L_l\rangle$ is the reduced dipole matrix between the fine states $|L_l\rangle$ and $|L_j\rangle$. The DLS between the two ground states is the subtraction of the ac Stark shifts.

In order to calculate the one-photon ac Stark shifts of the cesium ground states, $|n\text{P}_{1/2,3/2}\rangle$ with $n$ from 6 to 14 are taken into account. Due to relatively small two-photon frequency detuning, the calculation of the two-photon ac Stark shifts only needs to consider the corresponding hyperfine structure of the higher energy level. All the adopted hyperfine states, the transition wavelengths, and  the educed dipole matrix elements are shown in Tables \ref{tab1s} and \ref{tab2s}.

\begin{table}
\caption{\label{tab1s} The energies of relevant levels of Cs and the corresponding hyperfine constants. All the data are from \cite{Sansonetti2009} and the references therein. }
\begin{ruledtabular}
\begin{tabular}{cccc}
State& Energy & A & B \\
& (cm$^{-1}$)& (MHz) & (MHz)\\
\hline
6S$_{1/2}$ & 0 & 2298.1579425 & \\
6P$_{1/2}$ & 11178.26815870 & 291.9309 & \\
6P$_{3/2}$ & 11732.3071041 & 50.28825 & $-0.494$ \\
5D$_{3/2}$ & 14499.2568 & 48.78 & 0.1 \\
5D$_{5/2}$ & 14596.84232 & $-21.24$ & 0.2 \\
7S$_{1/2}$ & 18535.5286 & 545.90 &  \\
7P$_{1/2}$ & 21765.348 & 94.35 &  \\
7P$_{3/2}$ & 21946.397 & 16.609 &  \\
6D$_{3/2}$ & 22588.8210 & 16.34 & $-0.1$ \\
6D$_{5/2}$ & 22631.6863 & $-4.66$ & 0.9 \\
8S$_{1/2}$ & 24317.149400 & 219.12 &   \\
8P$_{1/2}$ & 25708.85473 & 42.97 &   \\
8P$_{3/2}$ & 25791.508 & 7.626 &   \\
7D$_{3/2}$ & 26047.8342 & 7.4 & $-0.1$ \\
7D$_{5/2}$ & 26068.7730 & $-1.7$ & 0.9 \\
9P$_{1/2}$ & 27636.9966 & 23.19 &   \\
9P$_{3/2}$ & 27681.6782 & 4.129 &   \\
10P$_{1/2}$ & 28726.8123 & 13.9 &   \\
10P$_{3/2}$ & 28753.6769 & 2.485 &   \\
11P$_{1/2}$ & 29403.42310 &   &   \\
11P$_{3/2}$ & 29420.824 & 1.6 &   \\
12P$_{1/2}$ & 29852.43153 &   &   \\
12P$_{3/2}$ & 29864.345 & 1.1 &   \\
13P$_{1/2}$ & 30165.66826 &   &   \\
13P$_{3/2}$ & 30174.178 & 0.77 &   \\
14P$_{1/2}$ & 30392.87183 &   &   \\
14P$_{3/2}$ & 30399.163 &   &  
\end{tabular}
\end{ruledtabular}
\end{table}

\begin{table}
\caption{\label{tab2s} The wavelengths ($\lambda$) and the related reduced matrices elements (D) adopted for the calculation of the magic conditions for cesium microwave transitions. The wavelengths are calculated from the energies shown in Table \ref{tab1s}. The reduced matrices elements marked by a) are experimental data summarized in \cite{Sansonetti2009}. The rest are theoretical results with b) from \cite{Arora2007} and c) from \cite{Safronova2016}.}
\begin{ruledtabular}
\begin{tabular}{ccc}
Transition & $\lambda$ & D \\
 & (nm) & ($\times e a_0$) \\
\hline
$6\text{S}_{1/2}\leftrightarrow 6\text{P}_{1/2}$ & 894.593 & 4.497$^a$ \\
$6\text{S}_{1/2}\leftrightarrow 6\text{P}_{3/2}$ & 852.347 & 6.331$^a$ \\
$6\text{S}_{1/2}\leftrightarrow 7\text{P}_{1/2}$ & 459.446 & 0.2755$^a$ \\
$6\text{S}_{1/2}\leftrightarrow 7\text{P}_{3/2}$ & 455.656 & 0.5862$^a$ \\
$6\text{S}_{1/2}\leftrightarrow 8\text{P}_{1/2}$ & 388.971 & 0.07227$^a$ \\
$6\text{S}_{1/2}\leftrightarrow 8\text{P}_{3/2}$ & 387.725 & 0.2107$^a$ \\
$6\text{S}_{1/2}\leftrightarrow 9\text{P}_{1/2}$ & 361.834 & 0.03229$^a$ \\
$6\text{S}_{1/2}\leftrightarrow 9\text{P}_{3/2}$ & 361.25 & 0.1154$^a$ \\
$6\text{S}_{1/2}\leftrightarrow 10\text{P}_{1/2}$ & 348.107 & 0.01624$^a$ \\
$6\text{S}_{1/2}\leftrightarrow 10\text{P}_{3/2}$ & 347.782 & 0.07216$^a$ \\
$6\text{S}_{1/2}\leftrightarrow 11\text{P}_{1/2}$ & 340.096 & 0.009573$^a$ \\
$6\text{S}_{1/2}\leftrightarrow 11\text{P}_{3/2}$ & 339.895 & 0.05290$^a$ \\
$6\text{S}_{1/2}\leftrightarrow 12\text{P}_{1/2}$ & 334.981 & 0.006271$^a$ \\
$6\text{S}_{1/2}\leftrightarrow 12\text{P}_{3/2}$ & 334.847 & 0.03983$^a$ \\
$6\text{S}_{1/2}\leftrightarrow 13\text{P}_{1/2}$ & 331.503 & 0.004228$^a$ \\
$6\text{S}_{1/2}\leftrightarrow 13\text{P}_{3/2}$ & 331.409 & 0.03228$^a$ \\
$6\text{S}_{1/2}\leftrightarrow 14\text{P}_{1/2}$ & 329.025 & 0.002977$^a$ \\
$6\text{S}_{1/2}\leftrightarrow 14\text{P}_{3/2}$ & 328.956 & 0.02506$^a$ \\
$6\text{P}_{1/2}\leftrightarrow 5\text{D}_{3/2}$ & 3011.15 & 7.015$^a$ \\
$6\text{P}_{3/2}\leftrightarrow 5\text{D}_{3/2}$ & 3490.97 & 9.919$^a$ \\
$6\text{P}_{3/2}\leftrightarrow 5\text{D}_{5/2}$ & 3614.09 & 3.1588$^a$ \\
$6\text{P}_{1/2}\leftrightarrow 7\text{S}_{1/2}$ & 1359.2 & 4.236$^b$ \\
$6\text{P}_{1/2}\leftrightarrow 8\text{S}_{3/2}$ & 761.1 & 1.026$^b$ \\
$6\text{P}_{1/2}\leftrightarrow 6\text{D}_{3/2}$ & 876.382 & 4.25$^b$ \\
$6\text{P}_{1/2}\leftrightarrow 7\text{D}_{3/2}$ & 672.515 & 2.05$^b$ \\
$6\text{P}_{3/2}\leftrightarrow 7\text{S}_{1/2}$ & 1469.89 & 6.47$^b$ \\
$6\text{P}_{3/2}\leftrightarrow 8\text{S}_{1/2}$ & 794.607 & 1.461$^b$ \\
$6\text{P}_{3/2}\leftrightarrow 6\text{D}_{3/2}$ & 921.106 & 2.10$^b$ \\
$6\text{P}_{3/2}\leftrightarrow 6\text{D}_{5/2}$ & 917.483 & 6.15$^b$ \\
$6\text{P}_{3/2}\leftrightarrow 7\text{D}_{3/2}$ & 698.542 & 0.976$^b$ \\
$6\text{P}_{3/2}\leftrightarrow 7\text{D}_{5/2}$ & 697.522 & 2.89$^b$ \\
$7\text{P}_{1/2}\leftrightarrow 8\text{S}_{1/2}$ & 3918.8 & 9.31$^c$ \\
$7\text{P}_{1/2}\leftrightarrow 6\text{D}_{3/2}$ & 12143.7 & 17.99$^c$ \\
$7\text{P}_{1/2}\leftrightarrow 7\text{D}_{3/2}$ & 2335.09 & 6.6$^c$ \\
$7\text{P}_{3/2}\leftrightarrow 8\text{S}_{1/2}$ & 4218.07 & 14.07$^c$ \\
$7\text{P}_{3/2}\leftrightarrow 6\text{D}_{3/2}$ & 15566 & 8.07$^c$ \\
$7\text{P}_{3/2}\leftrightarrow 6\text{D}_{5/2}$ & 14592.4 & 24.35$^c$ \\
$7\text{P}_{3/2}\leftrightarrow 7\text{D}_{3/2}$ & 2438.17 & 3.3$^c$ \\
$7\text{P}_{3/2}\leftrightarrow 7\text{D}_{5/2}$ & 2425.79 & 9.6$^c$ \\
$8\text{P}_{1/2}\leftrightarrow 7\text{D}_{3/2}$ & 29500.3 & 32$^c$ \\
$8\text{P}_{3/2}\leftrightarrow 7\text{D}_{3/2}$ & 39012.8 & 14.35$^c$ \\
$8\text{P}_{3/2}\leftrightarrow 7\text{D}_{5/2}$ & 36066.6 & 43.2$^c$ \\
\end{tabular}
\end{ruledtabular}
\end{table}

Same procedures are repeated to calculate the magic conditions of microwave transitions of rubidium ($^{87}$Rb and $^{89}$Rb). In the calculation of the one-photon ac Stark shifts of ground states, the states $|n\text{P}_{1/2,3/2}\rangle$ with the principal quantum numbers $n$ from 5 to 13 are taken into account. The hyperfine states, transition wavelengths, and the reduced dipole matrix elements used for the calculation of the magic conditions are summarized in Tables \ref{tab3s} and \ref{tab4s}.

\begin{table*}
\caption{\label{tab3s} The energies of the relevant levels of Rb and the corresponding hyperfine constants. All the data are from \cite{Sansonetti2006} and the references therein except a) from \cite{Lee2015} and b) from \cite{Moon2009}. }
\begin{ruledtabular}
\begin{tabular}{cccccc}
State& Energy & A$^{87}$ & B$^{87}$ & A$^{85}$ & B$^{85}$ \\
& (cm$^{-1}$)& (MHz) & (MHz) & (MHz) & (MHz) \\
\hline
5S$_{1/2}$ & 0 & 3417.34130545 & & 1011.91109852 & \\
5P$_{1/2}$ & 12578.950 & 409.2 & & 120.7 & \\
5P$_{3/2}$ & 12816.545 &84.7184 & 12.496 & 25.04 & 26.01\\
4D$_{5/2}$ & 19355.203 & $-16.779^a$ & 4.1120$^a$ & $-5.0080^a$ & 7.150$^a$\\
4D$_{3/2}$ & 19355.649 & 24.8$^b$ & 2.19$^b$ & 7.31$^b$ & 4.53$^b$\\
6S$_{1/2}$ & 20132.510 & 807.66 &  & 239.180 & \\
6P$_{1/2}$ & 23715.081 & 132.56 &  & 39.12 & \\
6P$_{3/2}$ & 23792.591 & 27.70 & 3.96 & 8.163 & 8.190 \\
7P$_{1/2}$ & 27835.02 & 59.93 &  & 17.7 & \\
7P$_{3/2}$ & 27870.11 & 12.57 & 1.71 & 3.711 & 3.69\\
6D$_{3/2}$ & 28687.127 & 2.3 & 1.6 &  &  \\
6D$_{5/2}$ & 28689.390 &  &  &  &  \\
8P$_{1/2}$ & 29834.94 &  &  &  &  \\
8P$_{3/2}$ & 29853.79 &  &  &  &  \\
9P$_{1/2}$ & 30958.91 &  &  &  &  \\
9P$_{3/2}$ & 30970.19 &  &  &  &  \\
10P$_{1/2}$ & 31653.85 &  &  &  &  \\
10P$_{3/2}$ & 31661.16 &  &  &  &  \\
11P$_{1/2}$ & 32113.55 &  &  &  &  \\
11P$_{3/2}$ & 32118.52 &  &  &  &  \\
12P$_{1/2}$ & 32433.50 &  &  &  &  \\
12P$_{3/2}$ & 32437.04 &  &  &  &  \\
13P$_{1/2}$ & 32665.03 &  &  &  & \\
13P$_{3/2}$ & 32667.63 &  &  &  & 

\end{tabular}
\end{ruledtabular}
\end{table*}

\begin{table}
\caption{\label{tab4s} The wavelengths ($\lambda$) and the related reduced matrices elements (D) adopted for the calculation of the magic conditions for rubidium microwave transitions. The wavelengths are calculated from the energies shown in Table \ref{tab3s}. The reduced matrices elements are deduced from the experimental values summarized in \cite{Sansonetti2006} except for the theoretical values with a) from \cite{Safronova2004}.}
\begin{ruledtabular}
\begin{tabular}{ccc}
Transition & $\lambda$ & D \\
 & (nm) & ($\times e a_0$) \\
\hline
$5\text{S}_{1/2}\leftrightarrow 5\text{P}_{1/2}$ & 794.979 & 4.231 \\
$5\text{S}_{1/2}\leftrightarrow 5\text{P}_{3/2}$ & 780.241 & 5.977 \\
$5\text{S}_{1/2}\leftrightarrow 6\text{P}_{1/2}$ & 421.673 & 0.1054 \\
$5\text{S}_{1/2}\leftrightarrow 6\text{P}_{3/2}$ & 420.299 & 0.5094 \\
$5\text{S}_{1/2}\leftrightarrow 7\text{P}_{1/2}$ & 359.26 & 0.1150 \\
$5\text{S}_{1/2}\leftrightarrow 7\text{P}_{3/2}$ & 358.807 & 0.1900 \\
$5\text{S}_{1/2}\leftrightarrow 8\text{P}_{1/2}$ & 335.177 & 0.05755 \\
$5\text{S}_{1/2}\leftrightarrow 8\text{P}_{3/2}$ & 334.966 & 0.1008 \\
$5\text{S}_{1/2}\leftrightarrow 9\text{P}_{1/2}$ & 323.009 & 0.03574 \\
$5\text{S}_{1/2}\leftrightarrow 9\text{P}_{3/2}$ & 322.891 & 0.06522 \\
$5\text{S}_{1/2}\leftrightarrow 10\text{P}_{1/2}$ & 315.917 & 0.02501 \\
$5\text{S}_{1/2}\leftrightarrow 10\text{P}_{3/2}$ & 315.844 & 0.04585 \\
$5\text{S}_{1/2}\leftrightarrow 11\text{P}_{1/2}$ & 311.395 & 0.01946 \\
$5\text{S}_{1/2}\leftrightarrow 11\text{P}_{3/2}$ & 311.347 & 0.03867 \\
$5\text{S}_{1/2}\leftrightarrow 12\text{P}_{1/2}$ & 308.323 & 0.01421 \\
$5\text{S}_{1/2}\leftrightarrow 12\text{P}_{3/2}$ & 308.29 & 0.02936 \\
$5\text{S}_{1/2}\leftrightarrow 13\text{P}_{1/2}$ & 306.138 & 0.01130 \\
$5\text{S}_{1/2}\leftrightarrow 13\text{P}_{3/2}$ & 306.113 & 0.02438 \\
$5\text{P}_{1/2}\leftrightarrow 4\text{D}_{3/2}$ & 1475.64 & 7.847$^a$ \\
$5\text{P}_{3/2}\leftrightarrow 4\text{D}_{5/2}$ & 1529.37 & 10.634$^a$ \\
$5\text{P}_{3/2}\leftrightarrow 4\text{D}_{3/2}$ & 1529.26 & 3.540$^a$ \\
$5\text{P}_{1/2}\leftrightarrow 6\text{S}_{1/2}$ & 1323.88 & 4.119$^a$ \\
$5\text{P}_{3/2}\leftrightarrow 6\text{S}_{1/2}$ & 1366.87 & 6.013$^b$ \\
$5\text{P}_{1/2}\leftrightarrow 6\text{D}_{3/2}$ & 620.803 & 1.18$^b$ \\
$5\text{P}_{3/2}\leftrightarrow 6\text{D}_{5/2}$ & 630.007 & 1.658$^b$ \\
$5\text{P}_{3/2}\leftrightarrow 6\text{D}_{3/2}$ & 630.097 & 0.558$^b$ \\
$6\text{P}_{1/2}\leftrightarrow 6\text{D}_{3/2}$ & 2011.24 & 1.989$^a$ \\
$6\text{P}_{3/2}\leftrightarrow 6\text{D}_{5/2}$ & 2042.15 & 2.974$^a$ \\
$6\text{P}_{3/2}\leftrightarrow 6\text{D}_{3/2}$ & 2013.09 & 1.012$^a$ \\
$7\text{P}_{1/2}\leftrightarrow 6\text{D}_{3/2}$ & 11735.6 & 31.422$^a$ \\
$7\text{P}_{3/2}\leftrightarrow 6\text{D}_{5/2}$ & 12205.8 & 42.481$^a$ \\
$7\text{P}_{3/2}\leftrightarrow 6\text{D}_{3/2}$ & 12239.6 & 14.161$^a$ \\
\end{tabular}
\end{ruledtabular}
\end{table}

\section{The triply magic conditions for rubidium }

The ODT magic conditions for rubidium can also be found by following the same procedure as for cesium. Here the magic conditions with 4D$_{3/2}$, 4D$_{5/2}$, 6S$_{1/2}$, 4D$_{3/2}$, and 4D$_{5/2}$ as the TPP excited states presented in Table \ref{tab5s}. The rest are not given due to the high Raman scattering rates and very short trapping wavelength. The magic conditions with 4D$_{3/2}$ and 4D$_{5/2}$ as the TPP excited states have wavelengths around 1033 nm. Although the fiber lasers are easy to match this wavelength, however, the trapping depth is may be too shallow to effectively trap the atoms in a MOT. The magic conditions with 6S$_{1/2}$ provide a deeper trap depth for $^{87}$Rb, and it can trap atoms with temperature on the level of hundreds of nK. However, the trap depth for $^{85}$Rb is much shallower in this case. The positive magic trap with 4D$_{3/2}$ and 4D$_{5/2}$ as the TPP excited states would be more attractive because of the arbitrarily high trap barrier. Similar to cesium, the magic conditions can be found for almost all the microwave transitions. Here, we only show the magic conditions for three experimental favorable transitions of (0,0), ($-1$,1) and (1,2) for $^{87}$Rb in Table \ref{tab5s}. The magic magnetic field for (0,0) is $B_0=0$ Gauss for both $^{87}$Rb and $^{85}$Rb. The corresponding residual second order coefficients are $k_M=1150$ Hz$/\text{Gauss}^2$ and $k_M=2590$ Hz$/\text{Gauss}^2$, respectively. The magic condition of magnetic field for ($-1$,1) is $B_0=3.23(1.21)$ Gauss and $k_M=862.7(2302)$ Hz$/\text{Gauss}^2$ for $^{87(85)}$Rb. 

\begin{table*}
\caption{\label{tab5s} Doubly magic frequency-intensity conditions of the ODT beam for rubidium-87 and rubidium-85 by 2-photon coupling to a higher energy level. GS means ground states with (m,n) representing ($|5S_{1/2},F=1(2),m_F=m\rangle$, $|5S_{1/2},F=2(3),m_F=n\rangle$). ES means the TPP excited state. $\lambda$, $\Delta \nu_0=\nu_0-\nu_\text{fine}/2$, and $I_0$ are the wavelength, relative frequency, and intensity of the trap beam at the magic point, respectively. $\nu_0$ and $\nu_\text{fine}$ are the laser frequency and frequency distance between the fine structures of ground state and higher excited state. $U_T$ is the trap potential. $k_\nu=\partial^2 \Delta^\text{(T)}_\text{DLS}/\partial \nu^2|_{(I_0, \nu_0)} $ and $k_I=\partial^2 \Delta^\text{(T)}_\text{DLS}/\partial I^2|_{(I_0, \nu_0)}$ are the residual coefficients for the second-order derivation of the total DLS over the trap beam frequency and intensity. $\Gamma^\text{(1p)}_\text{RS}$ and $\Gamma^\text{(2p)}_\text{S}$ are the 1-photon Raman scattering rate and 2-photon scattering rate, respectively. The wavelengths and transition matrices for the calculation are from \cite{Sansonetti2006} and \cite{Safronova2004}.}
\begin{ruledtabular}
\begin{tabular}{ccccccccccc}
GS & ES & $\lambda$& $\Delta \nu_0$ & $I_0$ & $U_T$ & $k_\nu$ & $k_I$ & $\Gamma^\text{(1p)}_\text{RS}$ &$\Gamma^\text{(2p)}_\text{S}$\\
& & (nm) & (GHz) & ($\text{GW}/\text{m}^2$)& ($\mu$K) & ($\times 10^{-18}$ Hz$^{-1}$) & (fHz$\cdot$ m$^4$/W$^2$) & (Hz) & (Hz)\\
\hline
\multicolumn{10}{c}{rubidium-87} \\
\hline
(0,0) & 4D$_{5/2}$ & 1033.31 & 0.429 & 0.0068 & $-0.87$ & 0.51 & 3.34 & $8.4\times 10^{-7}$ & $1.9\times 10^{-4}$ \\
(0,0) & 4D$_{3/2}$ & 1033.29 & 0.424 & 0.0090 & $-1.2$ & 0.68 & 2.50 & $1.1\times 10^{-6}$ & $4.1\times 10^{-5}$ \\
(0,0) & 6S$_{1/2}$ & 993.4 & 0.326 & 0.0439 & $-6.6$ & 7.66 & 0.68 & $1.1\times 10^{-5}$ & $2.9\times 10^{-3}$ \\
($-1$,1) & 6S$_{1/2}$ & 993.4 & 0.326 & 0.0431 & $-6.5$ & 7.38 & 0.68 & $1.1\times 10^{-5}$ & $2.7\times 10^{-3}$ \\
(1,2) & 6S$_{1/2}$ & 993.4 & 0.326 & 0.0419 & $-6.3$ & 6.97 & 0.68 & $1.0\times 10^{-5}$ & $2.6\times 10^{-3}$ \\
(0,0) & 6D$_{5/2}$ & 697.122 & 0.427 & 0.302 & 88.0 & 84.4 & 2.71 & $1.2\times 10^{-3}$ & $9.6\times 10^{-4}$ \\
(0,0) & 6D$_{3/2}$ & 697.177 & 0.427 & 0.599 & 175 & 168 & 1.36 & $2.5\times 10^{-3}$ & $3.2\times 10^{-4}$ \\
\hline
\multicolumn{10}{c}{rubidium-85} \\
\hline
(0,0) & 4D$_{5/2}$ & 1033.31 & 0.127 & 0.0013 & $-0.17$ & 0.23 & 75.1 & $5.7\times 10^{-8}$ & $3.7\times 10^{-5}$ \\
(0,0) & 4D$_{3/2}$ & 1033.29 & 0.126 & 0.0018 & $-0.23$ & 0.30 & 56.3 & $7.8\times 10^{-8}$ & $8.2\times 10^{-6}$ \\
(0,0) & 6S$_{1/2}$ & 993.4 & 0.097 & 0.0086 & $-1.31$ & 3.4 & 15.3 & $2.2\times 10^{-6}$ & $5.6\times 10^{-4}$ \\
(0,0) & 6D$_{5/2}$ & 697.122 & 0.126 & 0.0595 & 17.4 & 37.5 & 6.10 & $2.4\times 10^{-4}$ & $1.9\times 10^{-4}$ \\
(0,0) & 6D$_{3/2}$ & 697.177 & 0.126 & 0.119 & 34.7 & 74.2 & 3.06 & $4.9\times 10^{-4}$ & $6.2\times 10^{-5}$ \\
\end{tabular}
\end{ruledtabular}
\end{table*}

\section{Estimation of the coherence time due to the inhomogeneous dephasing }

In this section we will introduce the method to estimate the coherence time due to inhomogeneous dephasing associated with the atom motion in the trap. The method follows the procedure in reference \cite{Kuhr2005}.

A hot atom in a dipole trap has total energy of $E= E_\text{kin}+U$, where $E_\text{kin}$ is the kinetic energy, and $U$ is the potential energy. The probability density of the total energy $E$ obeys a three-dimensional Boltzmann distribution
\begin{equation}\label{eq8s}
p(E)=\frac{E^2}{2(k_\text{B} T)^3 } \exp{\left(-\frac{E}{k_\text{B} T}\right)},
\tag{S8}
\end{equation}
where $k_\text{B}$ is the Boltzmann constant, and $T$ is the temperature of the atoms. The average potential energy of the atom is half of the total enegy, $U_\text{avg}=E/2$. In a magic trap discussed in the main text of our paper, the first-order dependence of the total differential light shift over the light intensity is cancelled and only the second-order dependence is concerned. Since the trapping potential of an ODT is proportional to the light field intensity, the total differential light shift also depends on the average potential energy to the second order, which can be expressed as 
\begin{align}\label{eq9s}
\delta_\text{DLS}^\text{(T)}&=\delta_0+k_\text{E} U_\text{avg}^2 \tag{S9a}\\
 &=\delta_0+\frac{1}{4}k_\text{E} E^2,
\tag{S9b}
\end{align}
where $\delta_0$ is the differential light shift under the magic conditions and $k_\text{E}$ is the corresponding second-order coefficient. $k_\text{E}$ is connected to the residual second-order coefficients of the total DLS over intensity $k_\text{I}$ by
\begin{equation}\label{eq10s}
k_\text{E} = \frac{I_0^2}{U_{T}^2} k_\text{I},
\tag{S10}
\end{equation}
where $I_0$ is the magic intensity, and $U_{T}$ is the corresponding trap depth. For simplicity we define a constant
\begin{equation}\label{eq11s}
A=\frac{1}{\sqrt{k_\text{E}}} \frac{1}{k_\text{B} T}.
\tag{S11}
\end{equation}
By using the three-dimensional Boltzmann distribution Eq. (\ref{eq8s}), the probability  density of DLS can be deduced as
\begin{equation}\label{eq12s}
p(\delta_\text{DLS}^\text{(T)})=\frac{A^4}{12 } (\delta_\text{DLS}^\text{(T)}-\delta_0)\exp{\left(- A \sqrt{\delta_\text{DLS}^\text{(T)}-\delta_0} \right)}.
\tag{S12}
\end{equation}
Therefore, the expected Ramsey signal is 
\begin{equation}\label{eq13s}
w_\text{Ramsey}(t)=\int^{\infty}_{\delta_0} p(\delta_\text{DLS}^\text{(T)}) \cos[(\delta_\text{DLS}^\text{(T)}-\delta')t] \, \mathrm{d} \delta_\text{DLS}^\text{(T)}
\tag{S13}
\end{equation}
with $\delta'$ the overall DLS introduced from other sources, eg the intentionally introduced DLS offset on the driving field and DLS induced by magnetic field. If we let $x=\delta_\text{DLS}^\text{(T)}-\delta_0$, by substituting Eq. (\ref{eq12s}) into Eq. (\ref{eq13s}) we get
\begin{equation}\label{eq14s}
w_\text{Ramsey}(t)=\int^{\infty}_{0}  \frac{A^4}{12 } x \exp{\left(- A \sqrt{x} \right)}\cos[(x-\delta'')t] \, \mathrm{d} x,
\tag{S14}
\end{equation}
where $\delta''=\delta'-\delta_0$. To get the result of this integral, we divide it into two parts 
\begin{equation}\label{eq15s}
\alpha (t)=\int^{\infty}_{0}  \frac{A^4}{12 } x \exp{\left(- A \sqrt{x} \right)}\cos{(xt)} \, \mathrm{d} x
\tag{S15}
\end{equation}
and
\begin{equation}\label{eq16s}
\beta (t)=\int^{\infty}_{0}  \frac{A^4}{12 } x \exp{\left(- A \sqrt{x} \right)} \sin{(xt)} \, \mathrm{d} x.
\tag{S16}
\end{equation}
The Ramsey interference signal is 
\begin{equation}\label{eq17s}
w_\text{Ramsey}(t)=\alpha (t) \cos{(\delta''t)} +\beta (t)\sin{(\delta''t)},
\tag{S17}
\end{equation} 
and the amplitude of the signal is 
\begin{equation}\label{eq18s}
A(t)=\sqrt{\alpha (t)^2 +\beta (t)^2}.
\tag{S18}
\end{equation} 
The coherence time is defined by the time duration $T_2'$: when $t=T_2'$ the the amplitude of the Ramsey interference signal drops to $1/e$.

\section{Enhancing the trap depth of the negative trap in doubly frequency-intensity conditions }

\begin{figure*}
\includegraphics[width=17 cm]{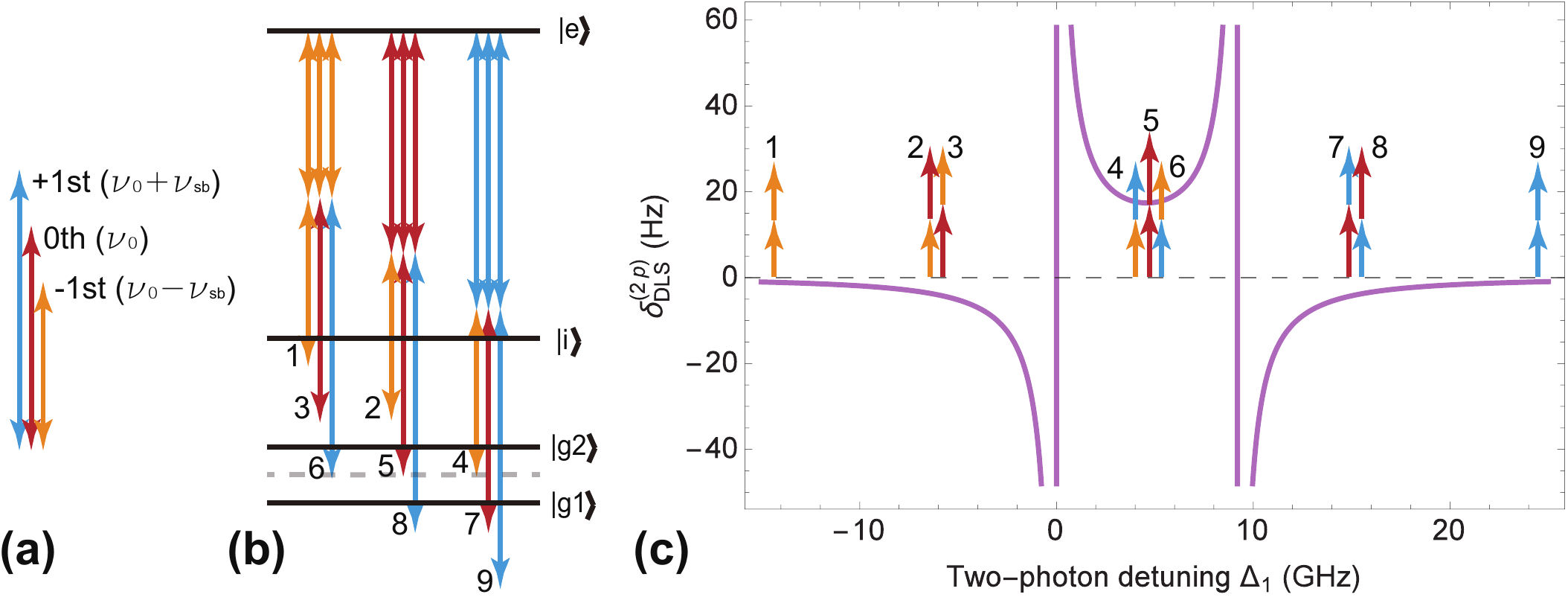}
\caption{\label{fig1s} (a) The sketch of photon energy from a phase modulated laser beam. The carrier (0th) photon has a frequency of $\nu_0$ and $\pm 1$-sideband photon has a frequency of $\nu_0 \pm \nu_\text{sb}$. The lengths of the arrows represent the photon energy. (b) Atom energy scheme with the possible TPP in the phase modulated ODT. (c) TPP-induced DLS calculated by Eq. (2) in the main text with $\delta_\text{hpf}=9.2$ GHz and $\frac{ \Omega_1^2\Omega_2^2}{4 \Delta^2}=40$. The arrow pairs show the possible TPP and the horizontal positions give the corresponding two-photon frequency detunings. In (b) and (c), each TPP is marked with a number from 1 to 9.}
\end{figure*}

\begin{table*}
\caption{\label{tab6s} Magic conditions of ODT beam for cesium by using a frequency modulated polychromatic ODT beam. $\nu_\text{sb}$ is the modulation frequency. The modulation depth is kept as 1.44, by which the power proportions of 0-th, 1st, and 2nd sidebands are 0.297, 0.301, and 0.047, respectively. Other symbols have same meaning as those in Table \ref{tab5s}. Only the microwave transition between clock states (0,0) are considered here.}
\begin{ruledtabular}
\begin{tabular}{ccccccccccc}
$\nu_\text{sb}$& ES & $\lambda$& $\Delta \nu_0$ & $I_0$ & $U_T$ & $k_\nu$ & $k_I$ & $\Gamma^\text{(1p)}_\text{RS}$ &$\Gamma^\text{(2p)}_\text{S}$\\
(GHz)& & (nm) & (GHz) & ($\text{GW}/\text{m}^2$)& ($\mu$K) & ($\times 10^{-18}$ Hz$^{-1}$) & (fHz$\cdot$ m$^4$/W$^2$) & (Hz) & (Hz)\\
\hline
9.4 & 5D$_{3/2}$ & 1379 & 0.281 & 0.211 & $-19.7$ & 4.45 & 0.81 & $2.3\times 10^{-5}$ & $1.1\times 10^{-4}$ \\
9.4 & 5D$_{5/2}$ & 1370 & 0.290 & 0.153 & $-14.5$ & 3.76 & 1.15 & $1.7\times 10^{-5}$ & $5.8\times 10^{-4}$ \\
7.0 & 7S$_{1/2}$ & 1079 & 0.219 & 1.14 & $-232$ & 195 & 0.57 & $3.3\times 10^{-3}$ & 0.37 \\
\end{tabular}
\end{ruledtabular}
\end{table*}

The trap depth for the magic conditions of some red (negative) traps for cesium, for example the state 5D coupled 1379-nm and 1370-nm ODT in Table I of main text, might be too shallow to directly load atoms from a laser-cooled atomic ensemble. To directly load atoms, one needs a deeper trap, but the magic conditions will be destroyed. This dilemma can be resolved by using a polychromatic trap beam. The idea is based on the fact that the TPP is red- and blue-detuned to the transitions of $|g1\rangle \leftrightarrow |e\rangle$ and $|g2\rangle \leftrightarrow |e\rangle$ simutaneously in our magic conditions. The TPP-induced DLS between $|g2\rangle$ and $|g1\rangle$ is then positive, which is used to compensate the negative OPP-induced DLS. If we consider that the trap laser beam is phase modulated with a frequency $\nu_\text{sb}$, so that the TPP caused by one carrier photon and one sideband photon is blue- or red-detuned to both of the two transitions (the TPPs with number 2, 3, 7, and 8 in Fig. \ref{fig1s}). The TPP-induced DLS will be negative. However, the TPP-induced DLS by two carrier photons (the TPP with number 5 in Fig. \ref{fig1s}) are still positive. Figure \ref{fig1s} shows the effect in detail. In fact, if only the carrier and $\pm1$-sideband photons are considered, the TPPs marked with 1--3 and 7--8 will induce a negative DLS and TPPs marked with 3--4 will induce positive DLS [see Fig. \ref{fig1s} (b) and (c)]. The overall effect is that the amount of TPP-induced DLS is decreased with the same ODT power in this polychromatic configuration, thus a higher ODT intensity is required to get the magic point again.  

To calculate the TPP-induced DLS in the polychromatic trap, it should be noted that there are two coupling configurations for TPP with two different photons. In these two configurations the lower part of transition $|g1(2)\rangle \leftrightarrow |i\rangle$ is coupled by either one of the two photons with different one-photon detuning, for example the TPPs marked by numbers 2 and 3, 4 and 6, 7 and 8 in Fig. \ref{fig1s}(b) and (c). For these TPP coupled by photons with unequal frequencies, the selection rules of $\Delta F=0$ and $\Delta m_F=0$ no longer apply. At this time Eq. (\ref{eq6s}) is invalid and the TPP induced DLS should be calculated by Eqs. (\ref{eq4s}), (\ref{eq5s}), and (\ref{eq7s}). However, Eq. (\ref{eq6s}) is still valid for the degenerated TPP (TPPs marked by numbers 1, 5, and 9). The DLS due to different TPP needs to be calculated separately and summed together to get the total DLS. Table \ref{tab6s} gives some of the magic frequency-intensity conditions of cesium clock states (0,0) in polychromatic traps. We can see that the main expenses for these deeper magic ODTs are the higher one- and two-photon scattering rates.

\begin{figure*}
\includegraphics[width=17 cm]{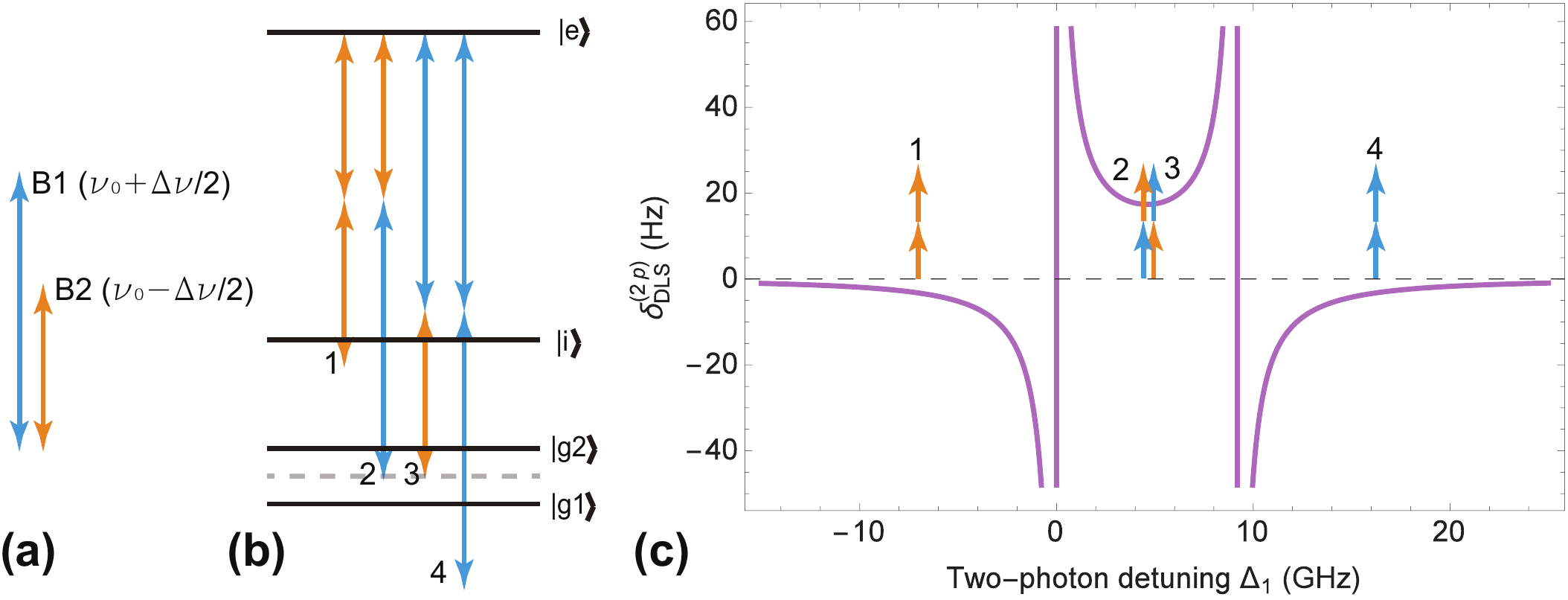}
\caption{\label{fig2s} (a) The sketch of photon energy from two laser beams B1 and B2 with frequencies of $\nu_0 + \Delta \nu /2$ and $\nu_0 - \Delta \nu /2$. The two beam can make a crossed dipole trap. The lengths of the arrows represent the photon energies. (b) Atom energy scheme with the possible TPP in the crossed dipole trap. (c) TPP-induced DLS calculated by Eq. (2) in the main text with $\delta_\text{hpf}=9.2$ GHz and $\frac{ \Omega_1^2\Omega_2^2}{4 \Delta^2}=40$. The arrow pairs show the possible TPP and the horizontal positions give the corresponding two-photon frequency detunings. In (b) and (c) each TPP is marked with number 1--4.}
\end{figure*}

The corresponding magic conditions of ODT beam for rubidium by using a frequency modulated polychromatic ODT beam are given in Table \ref{tab7s}.

\begin{table*}
\caption{\label{tab7s} Magic conditions for rubidium by using a frequency modulated polychromatic ODT beam. $\nu_\text{sb}$ is the modulation frequency and the modulation depth is kept as 1.44, by which the power proportions of 0-th, 1st, and 2nd sidebands are 0.297, 0.301, and 0.047. Other symbols have the same meaning as those in Tables \ref{tab5s}. Only the microwave transition between clock states (0,0) are considered here.}
\begin{ruledtabular}
\begin{tabular}{ccccccccccc}
$\nu_\text{sb}$& ES & $\lambda$& $\Delta \nu_0$ & $I_0$ & $U_T$ & $k_\nu$ & $k_I$ & $\Gamma^\text{(1p)}_\text{RS}$ &$\Gamma^\text{(2p)}_\text{S}$\\
(GHz)& & (nm) & (GHz) & ($\text{GW}/\text{m}^2$)& ($\mu$K) & ($\times 10^{-18}$ Hz$^{-1}$) & (fHz$\cdot$ m$^4$/W$^2$) & (Hz) & (Hz)\\
\hline
\multicolumn{10}{c}{rubidium-87} \\
\hline
6.9 & 4D$_{5/2}$ & 1033.31 & 0.430 & 0.0704 & $-9.1$ & 4.12 & 3.20 & $9.0\times 10^{-6}$ & $6.4\times 10^{-3}$ \\
6.9 & 4D$_{3/2}$ & 1033.29 & 0.424 & 0.094 & $-12.2$ & 5.51 & 2.40 & $1.2\times 10^{-5}$ & $1.4\times 10^{-3}$ \\
5.3 & 6S$_{1/2}$ & 993.4 & 0.326 & 0.438 & $-66.1$ & 66.6 & 0.679 & $1.1\times 10^{-4}$ & $8.7\times 10^{-2}$ \\
\hline
\multicolumn{10}{c}{rubidium-85} \\
\hline
3.0 & 4D$_{5/2}$ & 1033.31 & 0.127 & 0.0146 & $-1.9$ & 1.62 & 6.83 & $6.4\times 10^{-6}$ & $1.4\times 10^{-3}$ \\
3.0 & 4D$_{3/2}$ & 1033.29 & 0.126 & 0.0195 & $-2.5$ & 2.17 & 5.12 & $8.1\times 10^{-7}$ & $3.2\times 10^{-4}$ \\
2.3 & 6S$_{1/2}$ & 993.4 & 0.097 & 0.0909 & $-19.7$ & 27.4 & 1.45 & $1.2\times 10^{-5}$ & $2.0\times 10^{-2}$ \\
\end{tabular}
\end{ruledtabular}
\end{table*}

Our polychromatic ODT scheme can also be used to built a crossed trap, where the beams must be at different optical frequencies to avoid making an optical lattice. It is also possible to find the doubly frequency-intensity condition and the triply magic conditions. The configurations of laser frequencies and the energy scheme are shown in Fig. \ref{fig2s}(a) and (b). The two laser beams B1 and B2 have frequencies $\nu_0 + \Delta \nu /2$ and $\nu_0 - \Delta \nu /2$, respectively, where $\nu_0 $ is the magic frequency. The frequency difference $\Delta \nu$ should be larger enough to promise that the TPP-induced DLS due to photons from the same beam have a negative value [TPP marked with 1 and 4 in Fig. \ref{fig2s}(c)] and that due to photons from different beams has a positive value [TPP marked with 2 and 3 in Fig. \ref{fig2s}(c)]. By this way, the crossed dipole trap would have larger depth and meet the magic frequency-intensity conditions simultaneously.

\end{document}